\documentstyle[twocolumn,epsf]{jpsj}

\catcode`\@=11

\def\simle{\mathrel{\mathpalette\@versim<}}   
\def\simge{\mathrel{\mathpalette\@versim>}}   
\def\@versim#1#2{\lower2.5pt\vbox{\baselineskip0pt \lineskip-.5pt
   \ialign{$\m@th#1\hfil##\hfil$\crcr#2\crcr\sim\crcr}}}

\newcommand{\bequ}{ \begin{equation} }
\newcommand{\eequ}{ \end{equation} }
\newcommand{\barr}{ \begin{array} }
\newcommand{\earr}{ \end{array} }
\newcommand{\beqarr}{ \begin{eqnarray} }
\newcommand{\eeqarr}{ \end{eqnarray} }

\newcommand{\baralpha}{ \begin{eqnal} \beqarr}
\newcommand{\earalpha}{ \eeqarr \end{eqnal}}

\catcode`\@=12

\def\runtitle{
}
\def\runauthor{Yukitoshi {\sc Motome} and Nobuo {\sc Furukawa}}

\def\sf{\rm}

\title{
Critical Temperature of Ferromagnetic Transition in\\
Three-Dimensional Double-Exchange Models
}

\author{
Yukitoshi {\sc Motome} and Nobuo {\sc Furukawa}$^{1}$ 
}
\inst{
Institute of Materials Science, University of Tsukuba,
Tsukuba, Ibaraki 305-0006 \\
$^{1}$Department of Physics, Aoyama Gakuin University, 
Setagaya, Tokyo 157-8572
}
\recdate{
}

\abst{
Ferromagnetic transition in three-dimensional double-exchange models
is studied by the Monte Carlo method.
Critical temperature $T_{\rm c}$ is precisely determined
by finite-size scaling analysis.
Strong spin fluctuations in this itinerant system significantly
reduce $T_{\rm c}$ from mean-field estimates.
By choosing appropriate parameters,
obtained values of $T_{\rm c}$ quantitatively agree with
experiments for the ferromagnetic metal regime of (La,Sr)MnO$_{3}$,
which is a typical perovskite manganite showing colossal magnetoresistance.
This indicates that the double-exchange mechanism alone
is sufficient to explain $T_{\rm c}$ in this material.
Critical exponents are also discussed.
}

\kword{
double-exchange model, colossal magnetoresistance manganites,
ferromagnetic transition, critical temperature, finite-size scaling,
Monte Carlo method
}

\begin{document}
\sloppy
\maketitle


Colossal magnetoresistance near the ferromagnetic transition
has attracted much attention to the strong interplay
between transport and magnetism in perovskite manganese oxides.
\cite{Ramirez1997, Furukawa1999}
This feature has been attributed to the double-exchange (DE) mechanism
\cite{Zener1951}
in the Hamiltonian
\begin{equation}
\label{eq:DEmodel}
{\cal H} = - \sum_{ij} \sum_{\sigma = \uparrow, \downarrow} t_{ij}
( c_{i \sigma}^{\dagger} c_{j \sigma} + {\rm h.c.} )
- J_{\rm H} \sum_{i} \mib{\sigma}_{i} \cdot \mib{S}_{i},
\end{equation}
where $c_{i \sigma}$ ($c_{i \sigma}^{\dagger}$) annihilates (creates)
a $\sigma$-spin electron at site $i$, and
$J_{\rm H}$ is the Hund's-rule coupling
between itinerant $e_{g}$ electrons with $\mib{\sigma}_{i}$ spin
and localized $t_{2g}$ moments $\mib{S}_{i}$.

Recently this simple DE scenario has been reexamined
from the viewpoint of whether model (\ref{eq:DEmodel})
quantitatively describes thermodynamics
in a typical material La$_{1-x}$Sr$_{x}$MnO$_{3}$ (LSMO),
or whether another mechanism is necessary.
Based on mean-field (MF) arguments,
Millis {\it et al.}
\cite{Millis1995}
claimed that the DE mechanism alone is insufficient
to reproduce experiments.
In particular, their estimate for the critical temperature $T_{\rm c}$
is one order of magnitude higher than experimental results.
On the other hand, the validity of the DE scenario
has been claimed by dynamical mean-field (DMF) studies.
\cite{Furukawa1999}
Magnetism as well as transport and optical properties in the LSMO compound
is well reproduced by model (\ref{eq:DEmodel}).

The DMF method includes dynamical fluctuations
which were completely neglected in the MF study.
DMF results suggest the importance of the fluctuations
in this itinerant system.
At the same time, the DMF method is not sufficient on its own
since it takes into account only local fluctuations.
In order to settle the controversy of whether the DE mechanism
alone explains the thermodynamics of LSMO,
it is necessary to make calculations
which include all fluctuation effects.

A numerical study is one of the promising candidates
for including the fluctuation effects appropriately.
Many surveys based on the Monte Carlo (MC) method have been performed.
However, their results have not yet converged
mainly due to finite-size effects.
Table \ref{table:Tc} summarizes the values of $T_{\rm c}$
from the MC studies as well as by other approximations.
MC calculations are performed for the system on finite-size clusters.
In order to discuss critical phenomena,
it is necessary to apply finite-size scaling analysis.
However, such a systematic study has not been performed thus far
since it has been difficult
to study large-size clusters in three dimensions (3D)
due to the rapid increase of the cpu time with the system size.

In this work we investigate the ferromagnetic transition
in model (\ref{eq:DEmodel})
by finite-size scaling analysis of MC results.
A new MC algorithm
\cite{Motome1999}
is applied to reduce the cpu time markedly,
which enables us to study larger clusters than ever.
Our results elucidate effects of thermal and spatial fluctuations quantitatively.
Our estimate for $T_{\rm c}$ agrees with those for LSMO
in the ferromagnetic metal regime
when parameters are chosen appropriately.
Critical exponents are also examined in comparison with experimental results.


Model (\ref{eq:DEmodel}) is studied
with a classical treatment of the localized moments $\mib{S}$,
as in other surveys listed in Table \ref{table:Tc}.
Then the fermion Hamiltonian defined for a given configuration
of $\{\mib{S}_{i}\}$ becomes quadratic in the fermion operators
and can be diagonalized with the matrix size proportional to the system size.
Configurations of $\{\mib{S}_{i}\}$ are stochastically sampled
in a MC fashion.
The bottleneck of the standard MC technique
\cite{Yunoki1998}
is the cost of the cpu time required to diagonalize
the fermion Hamiltonian in each MC sample.
The cost of cpu time is of the order of $N^{3}$
where $N$ is the system size.
We apply a new algorithm
\cite{Motome1999}
which replaces the exact diagonalization
by a moment expansion of the density of states.
\cite{Wang1994, Silver1994}
Calculations of the moments can be easily performed on parallel computers.
The cpu time is reduced to the order of $(N^{2} \log N) / N_{\rm PE}$
where $N_{\rm PE}$ is the number of nodes in the parallel calculation.
For details of the method, readers are referred to ref.
\citen{Motome1999}.

\vspace{-4mm}

\newcommand{\lwA}[1]{\smash{\lower1.6ex\hbox{#1}}}
\newcommand{\lwB}[1]{\smash{\lower5.0ex\hbox{#1}}}

\begin{table}[ht]
\caption{
The critical temperature $T_{\rm c}$ of ferromagnetic transition
in the DE model (\ref{eq:DEmodel}) estimated by various methods.
(MF: mean-field approximation,
HTE: high-temperature expansion,
DMF: dynamical mean-field theory,
MC: Monte Carlo calculation)
The results are for $x=0.5$ and $J_{\rm H}=\infty$
except for noted cases.
Errors in the last digit are shown in parentheses.
$W$ is the half bandwidth of noninteracting electrons.
The third column is $T_{\rm c}$ when $W$ is taken as $1$ eV.
The experimental result (EXP) is also denoted for comparison.
See text for details.
}
\begin{tabular}{@{\hspace{\tabcolsep}\extracolsep{\fill}}cllc}
\noalign{\hrule height0.8pt}
method & T$_{\rm c}/W$ & T$_{\rm c}$ (K) & Ref.\\
\noalign{\hrule height0.8pt}
MF & 0.3$^{\rm a}$ & 3500$^{\rm a}$ & \citen{Millis1995}) \\
\hline
HTE & 0.027 & 310 & \citen{Roder1997}) \\
\hline
\lwA{DMF}
 & 0.0809(3) & 942(4) & \lwA{\citen{Furukawa1995})} \\
 & 0.0714(1)$^{\rm b}$ & 831(1)$^{\rm b}$ & \\
\hline
MC & 0.018(1) & 210(10) & \citen{Yunoki1998}) \\
MC & 0.0313(8) & 364(9) & \citen{Calderon1998}) \\
MC & 0.0208 & 242 & \citen{Yi2000}) \\
\hline
\lwB{MC} & 0.037(3)$^{\rm c}$ & 430(40)$^{\rm c}$ & \\
 & 0.0362(8)$^{\rm d}$ & 422(9)$^{\rm d}$ & present \\
 & 0.022(3)$^{\rm b,c}$ & 260(40)$^{\rm b,c}$ & work \\
 & 0.0210(7)$^{\rm b,d}$ & 245(8)$^{\rm b,d}$ & \\
\hline
EXP & & 369$^{\rm e}$ & \citen{Tokura1994}) \\
\noalign{\hrule height0.8pt}
\multicolumn{4}{l}{a. $x=0.3$ and $J_{\rm H}=\infty$,
\ \ b. $x=0.5$ and $J_{\rm H}=4W$} \\
\multicolumn{4}{l}{c. without assumption for the exponent $\beta$} \\
\multicolumn{4}{l}{d. with assumption $\beta=0.365$}\\
\multicolumn{4}{l}{e. $x=0.3$}
\end{tabular}
\label{table:Tc}
\end{table}

\vspace{-6mm}

We consider the hopping $t_{ij} = t$ only between nearest-neighbor sites
on a 3D cubic lattice.
In the following we study the ferromagnetic transition
for two parameters, $J_{\rm H} = \infty$ and $4W$,
where $W$ is the half bandwidth of noninteracting electrons
($W = 6t$ in 3D).
For the case of $J_{\rm H} = \infty$,
components of conduction electrons with spins antiparallel to
localized moments are projected out, and therefore
model (\ref{eq:DEmodel}) simply takes the spinless form of
${\cal H} = - \tilde{t} \sum_{ij}
(c_{i}^{\dagger} c_{j} + {\rm h.c.})$,
where
$\tilde{t} = t [ \cos (\theta_{i}/2) \cos (\theta_{j}/2)
+ \sin (\theta_{i}/2) \sin (\theta_{j}/2)
e^{{\rm i} (\phi_{j} - \phi_{i})} ]$.
\cite{Anderson1955}
The matrix dimension of the Hamiltonian is halved
and the cpu time is reduced.
Moreover, the moment expansion converges rapidly
due to the simple band structure.
In the following calculations,
we perform the moment expansion up to the order of
$M=20$ for $J_{\rm H} = \infty$ and $M=40$ for $J_{\rm H} = 4W$
in order to ensure sufficient precision for MC updates.
\cite{Motome1999}
For both values of $J_{\rm H}$,
we control the chemical potential so that the electron density
$x = \langle \sum_{i \sigma} c_{i \sigma}^{\dagger} c_{j \sigma} \rangle$
becomes $0.5$.
Here, the bracket denotes the thermal average for the grand canonical ensemble.
The critical temperature $T_{\rm c}$ in model (\ref{eq:DEmodel})
is known to have a maximum at $x \simeq 0.5$.
A closed-shell condition is necessary to obtain
full polarization in the ground state ($T=0$).
We chose the system size $N = N_{x} \times N_{y} \times N_{z}$
as $4 \times 4 \times 4$, $6 \times 6 \times 4$,
$6 \times 6 \times 8$ and $8 \times 8 \times 8$
under the periodic boundary conditions for the $x$ and $y$ directions
and the antiperiodic boundary condition for the $z$ direction.
We have typically run $10,000$ MC samplings for measurements
after $1,000$ MC steps for thermalization.
Measurement is divided into five bins to estimate
the statistical error by the variance among the bins.

We determine the critical temperature $T_{\rm c}$ by two quantities.
One is the Binder parameter
\cite{Binder1981}
defined by
\begin{equation}
g = 1 - \langle \mib{q}^{4} \rangle / 3 \langle \mib{q}^{2} \rangle^{2},
\end{equation}
where $\mib{q}$ is the order parameter which is taken as
$\sum_{i} \mib{S}_{i}$ in this case.
For $T > T_{\rm c}$ ($T < T_{\rm c}$),
the Binder parameter decreases (increases) as the system size $N$ increases.
Thus $T_{\rm c}$ is determined by a crossing point of $g$ for various $N$.
The other quantity is the magnetization defined by
\begin{equation}
m = [ S(\mib{k}=0) / N ]^{1/2},
\end{equation}
where the spin structure factor $S(\mib{k})$ is given by
\begin{equation}
\label{eq:S(k)}
S(\mib{k}) = \sum_{ij} \langle \mib{S}_{i} \mib{S}_{j} \rangle
e^{{\rm i} \mib{k} \cdot \mib{r}_{ij}} / N.
\end{equation}
The thermodynamic limit of $m$ is obtained
from the system-size extrapolation of $S(0)$.

\begin{figure}
\epsfxsize=6.5cm
\centerline{\epsfbox{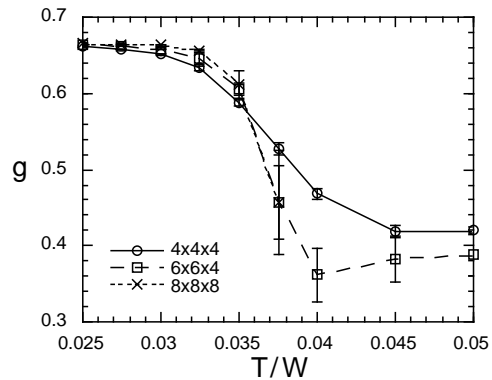}}
\caption{
Binder parameter for $J_{\rm H} = \infty$.
The lines are guides to the eye.
}
\label{fig:Binder JHInf}
\end{figure}

The case of $J_{\rm H} = \infty$ is discussed first.
Figure \ref{fig:Binder JHInf} shows the Binder parameter for this case.
The Binder parameters for different system sizes
cross at $T \sim 0.035W$
although it is difficult to determine $T_{\rm c}$ precisely
due to the statistical error bars.

\begin{figure}
\epsfxsize=6.5cm
\centerline{\epsfbox{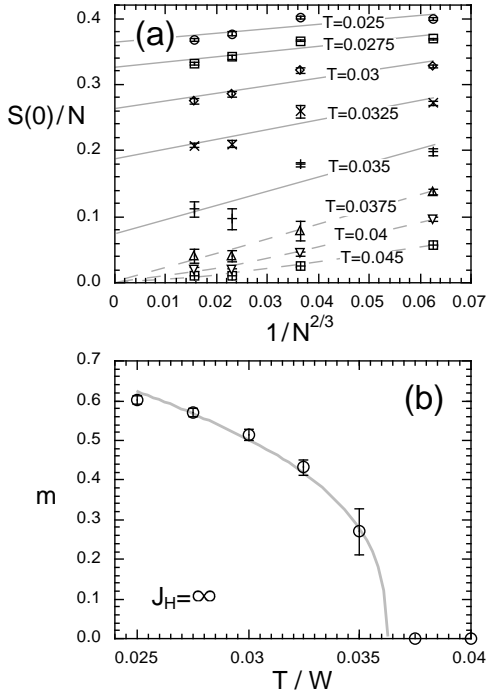}}
\caption{
(a) System-size extrapolation of the spin structure factor and
(b) magnetization in the thermodynamic limit for $J_{\rm H} = \infty$.
The curve in the figure is the least-squares-fit 
to $m \propto (T-T_{\rm c})^\beta$. See text for details.
}
\label{fig:mag JHInf}
\end{figure}

We next study the magnetization.
Figure \ref{fig:mag JHInf}(a) displays the system-size extrapolation of $S(0)/N$.
The data for $T \le 0.035W$ scale well with $N^{-2/3}$, and
are extrapolated to finite values in the limit of $N = \infty$.
This scaling is consistent with $\mib{k}^{2}$-dependence
of the energy cutoff for magnons.
\cite{Furukawa1996}
The extrapolated values are
summarized as the magnetization in Fig. \ref{fig:mag JHInf}(b).
The onset of the finite magnetization $m$ agrees well
with the crossing point of the Binder parameter in Fig. \ref{fig:Binder JHInf}.
The estimate for $T_{\rm c}$ is $T_{\rm c} \simeq (0.037 \pm 0.003)W$.

\vspace{-4mm}

To determine the value of $T_{\rm c}$ precisely,
we fit the magnetization data
by assuming the scaling law of $m \propto (T_{\rm c}-T)^{\beta}$.
The value of the exponent $\beta$ has not been determined thus far.
At present, our data are insufficient to make estimates
for both $T_{\rm c}$ and $\beta$ simultaneously
within the desired accuracy.
Instead we assume the exponent in 3D Heisenberg models $\beta = 0.365$,
\cite{LeGuillou1977}
and determine the value of $T_{\rm c}$.
The data are well fitted, as shown in Fig. \ref{fig:mag JHInf} (b).
The fit gives $T_{\rm c} = (0.0362 \pm 0.0008)W$.

\begin{figure}
\epsfxsize=6.5cm
\centerline{\epsfbox{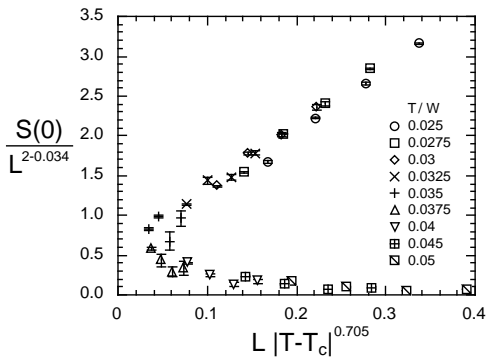}}
\caption{
Finite-size scaling for $J_{\rm H} = \infty$.
}
\label{fig:scaling}
\end{figure}

The assumption of $\beta$ in the universality class of 3D Heisenberg models
is tested by finite-size scaling analysis.
The scaling hypothesis gives the scaling relation
for the spin structure factor (\ref{eq:S(k)}) in the form
\begin{equation}
\label{eq:S(0)scaling}
S(0) = L^{2-\eta} f ( L |T-T_{\rm c}|^{\nu} ),
\end{equation}
where $L=N^{1/3}$ and $f$ is the scaling function.
The critical exponents are predicted to be
$\eta = 0.034$ and $\nu = 0.705$ in this universality class.
\cite{LeGuillou1977}
We plot $S(0) / L^{2-0.034}$
as a function of $L |T-T_{\rm c}|^{0.705}$ with $T_{\rm c} = 0.0362W$
in Fig. \ref{fig:scaling}.
The data appear to follow a universal function
(there are two branches for $T>T_{\rm c}$ and $T<T_{\rm c}$)
in this temperature range.
This is consistent with the assumption of $\beta=0.365$ in the fit
in Fig. \ref{fig:mag JHInf}(b).
Later we will discuss the critical exponents
in comparison with the experimental results.

\vspace{-7mm}

\vspace{-4mm}

\begin{figure}
\epsfxsize=6.5cm
\centerline{\epsfbox{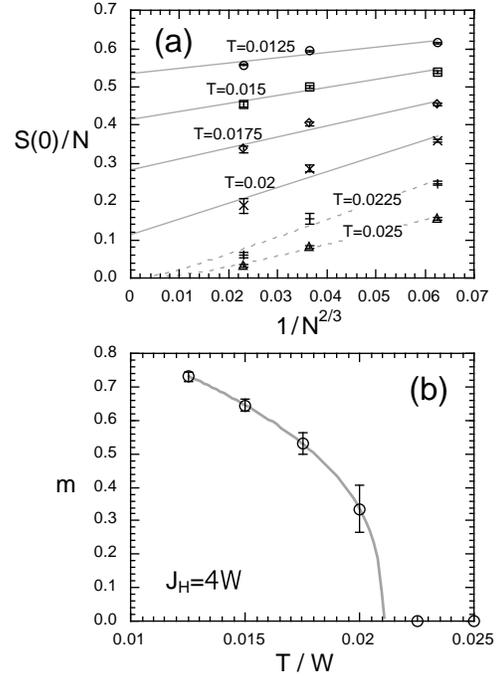}}
\caption{
(a) System-size extrapolation of the spin structure factor and
(b) magnetization in the thermodynamic limit for $J_{\rm H}=4W$.
The curve in the figure is the least-squares-fit 
to $m \propto (T-T_{\rm c})^\beta$. See text for details.
}
\label{fig:mag JH4}
\end{figure}

We now turn to the case of $J_{\rm H} = 4W$.
As in the case of $J_{\rm H}=\infty$,
an estimate for $T_{\rm c}$ is obtained more accurately
by the magnetization measurement.
Figure \ref{fig:mag JH4} shows
(a) the system-size extrapolation of $S(0)/N$ and
(b) the magnetization in the thermodynamic limit.
We have $T_{\rm c} \simeq (0.022 \pm 0.003)W$ from Fig. \ref{fig:mag JH4}(b).
If we assume the scaling $m \propto (T_{\rm c}-T)^{\beta}$
with the 3D Heisenberg exponent $\beta=0.365$
as in Fig. \ref{fig:mag JHInf}(b),
the fit for $T \le 0.02W$ gives $T_{\rm c} = (0.0210 \pm 0.0007)W$
in this case.

\vspace{-5mm}


We compare the results of $T_{\rm c}$ with the other theoretical estimates
listed in Table \ref{table:Tc}.
Our results are one order of magnitude lower than the MF estimates.
This elucidates the importance of the strong fluctuations in this system.
The MF approximation neglects all the fluctuations,
i.e., the thermal and spatial ones.
The DMF theory gives
an intermediate approximation in the sense that
it includes only the thermal fluctuation and neglects the spatial one.
Comparison among MF, DMF and our MC results
clarifies the fluctuation effects quantitatively:
the thermal fluctuation significantly suppresses $T_{\rm c}$
from $0.3W$ to $\sim 0.08W$, and moreover
the spatial fluctuation reduces $T_{\rm c}$ by more than half
from $\sim 0.08W$ to $\sim 0.04W$
in the case of $J_{\rm H}=\infty$.
Both fluctuations are important and
cannot be neglected in a discussion of the phase transition.

The previous MC results listed in Table \ref{table:Tc}
deviate considerably from our results.
These deviations arise from the finite-size effects and
the method used to determine the transition temperature.
$T_{\rm c}$ has been estimated by
spin correlation between the farthest sites,
\cite{Yunoki1998}
the second derivative of the magnetization
\cite{Calderon1998}
and the scaling fit for the data of a finite-size cluster.
\cite{Yi2000}
In order to discuss the phase transition in the strict sense,
these methods are uncontrolled in the absence of
system-size extrapolation.

The values of $T_{\rm c}$ are compared with the experimental result
in Table \ref{table:Tc}.
Here, we take the half bandwidth $W$ to be $1$ eV, which is typical
for $3d$ transition metal oxides.
The bandwidth $W=1$ eV is also consistent with band calculations.
\cite{Hamada1995, Pickett1997, Papaconstantopoulos1998}
Thus our results, $T_{\rm c} \simeq 430$ K for $J_{\rm H}=\infty$ and
$T_{\rm c} \simeq 260$ K for $J_{\rm H}=4W$, agree well with
the experimental values of $T_{\rm c}$
in the ferromagnetic metal regime, for instance,
$T_{\rm c} = 369$ K at $x=0.3$.
More quantitative agreement will be obtained
by tuning the value of $J_{\rm H}/W$
and the doping concentration $x$.
We conclude that
once the fluctuation effects are fully taken into account,
the DE mechanism gives a quantitative description
of $T_{\rm c}$ in LSMO as a major origin.

Our results strongly support a view of LSMO
in the ferromagnetic metal regime around $x=0.3$ as a canonical DE system,
which has been pointed out by the DMF studies.
\cite{Furukawa1999}
Note that the compound is far
from an antiferromagnetic/ferromagnetic insulator at $x \simle 0.15$
\cite{Urushibara1995}
and an $A$-type antiferromagnetic metal at $x \simge 0.5$.
\cite{Akimoto1998}
These instabilities are beyond the scope of model (\ref{eq:DEmodel}).
Detailed comparison between experiments involving this compound and
theories in model (\ref{eq:DEmodel}) provides a good starting point
to understanding not only the above instabilities
but physical properties in other related materials.
MC study of other physical properties in 3D
is a subject for further study.

In our analysis, the universality class of 3D Heisenberg models is assumed.
We also analyze the data by assuming
the universality class of the mean-field theory, i.e.,
$\beta=\nu=0.5$ and $\eta=0$.
In this case, the finite-size scaling plot (\ref{eq:S(0)scaling})
does not show good convergence
compared to the data in Fig. \ref{fig:scaling}.
Our data fit better with the 3D Heisenberg exponents
than do the mean-field ones.
However, the present data are not sufficiently precise
to either confirm the universality class 
of the DE model as that of the 3D Heisenberg model, or
to identify it as a new universality class
with similar exponents.

Concerning the experiments,
values of the critical exponents for LSMO remain controversial.
Estimates for $\beta$ are scattered
in the range of $0.3 \simle \beta \simle 0.5$.
\cite{Heffner1996, Martin1996, Lofland1997, Ghosh1998, Vasiliu-Doloc1998}
A recent study claims that
the estimate for the critical exponent may depend
on the range of the critical region assumed for fitting:
\cite{Schwartz2000}
$\beta \simeq 0.5$ is obtained for the relatively narrow critical region
$(T_{\rm c} - T)/T_{\rm c} \simle 0.1$
while $\beta \simeq 0.3$ is found for the relatively wide critical region
$(T_{\rm c} - T)/T_{\rm c} \simge 0.2$.
In our calculation, data of the magnetization are fitted
over the wide range of $(T_{\rm c} - T)/T_{\rm c} \simle 0.4$.
Considering the experimental situation,
we cannot exclude the possibility
that the results for the exponents in the DE model may be modified
when the system further approaches to the critical point.
Further investigation is required to determine
the value of the exponents with precision.


To summarize, we have investigated ferromagnetic transition
in the three-dimensional double-exchange model
by Monte Carlo calculations.
The new algorithm based on the moment expansion method
has been applied to study large clusters.
The critical temperature is determined precisely
by finite-size scaling analysis.
Comparison with other theoretical surveys quantitatively elucidates
the importance of thermal and spatial fluctuations in this itinerant system.
The critical temperature estimated with appropriate parameters
agrees well with experiments
in the ferromagnetic metal regime of (La,Sr)MnO$_{3}$.
The double-exchange mechanism alone is sufficient to reproduce
the ferromagnetic transition temperature in this material.
The critical exponent in our analysis is consistent
with experimental trends
although more detailed study is necessary
both theoretically and experimentally.


The authors thank H. Nakata for helpful support
in developing parallel-processing systems.
The computations have been performed mainly 
using the facilities in the AOYAMA+ project
(http://www.phys.aoyama.ac.jp/\~{}\\
aoyama+).
This work is supported by  ``a Grant-in-Aid from the Ministry of Education,
 Science, Sports and Culture''.


\end{document}